\documentclass[a4paper,11pt]{article}
\usepackage{pos}
 \usepackage[nice]{nicefrac}
\def\HP{\hphantom{\alpha}} 
\usepackage[utf8]{inputenc}
\usepackage{graphicx}
\usepackage{amsmath}
\usepackage{amsfonts}
\usepackage{physics}
\usepackage{xspace}
\usepackage{bm}
\usepackage{mathrsfs}
\usepackage{nicefrac}
\usepackage{multirow}
\usepackage[format=plain,justification=RaggedRight,labelsep=endash,font=small,labelfont=sc]{caption}
\usepackage{stackengine}

\usepackage[export]{adjustbox}
\newcommand{\FUG}{\ensuremath{\xi}}

\def\FUGM{\a_M}
\def\GLW{{\rm GLW}}
\newcommand{\Inqr}[3]{I_{#1#2}^{(#3)}}

\definecolor {darkgreen}{rgb}{0.2,0.7,0.2}

\def\be{\begin{equation}}
	\def\ee{\end{equation}}
\newcommand{\bel}[1]{\begin{eqnarray}\label{#1}}
	\newcommand{\eel}{\end{eqnarray}}
\def\barr{\begin{array}}
	\def\earr{\end{array}}
\def\beq{\begin{eqnarray}}
	\def\eeq{\end{eqnarray}}
\def\bfig{\begin{figure}}
	\def\efig{\end{figure}}
\def\lt{\left}
\def\rt{\right}
\def\CHI{\chi}
\newcommand{\nn}{\nonumber}
\newcommand{\f}[2]{\frac{#1}{#2}}

\newcommand{\p}{\partial}

\newcommand{\rf}[1]{Eq.~(\ref{#1})}

\newcommand{\rfn}[1]{(\ref{#1})}

\def\a{\alpha}
\def\b{\beta}
\def\g{\gamma}
\def\d{\delta} 
\def\r{\rho}
\def\s{\sigma}
\def\c{\chi}
 
\def\LR{\left(} 
\def\RR{\right)}
\def\LS{\left[} 
\def\RS{\right]}
\def\LC{\left{} 
\def\RC{\right}}
\def\LA{\left\langle}
\def\RA{\right\rangle}
\def\LD{\left.}
\def\RD{\right.}
\def\HP{\hphantom{\alpha}} 

\def\half{\frac{1}{2}}

\def\GLW{{\rm GLW}}
\def\LRFF{{\rm LRFF}}

\def\nU{n_{(0)}}
\def\eU{\varepsilon_{(0)}}
\def\PU{P_{(0)}}
\def\sU{s_{(0)}}

\def\nP{n_{}}
\def\eP{\varepsilon_{}}
\def\PP{P_{}}
\def\sP{s_{}}
\def\wP{w_{}}

\newcommand{\lab}[1]{\label{#1}}

\def\pmu{p^\mu}
\def\pnu{p^\nu}

\def\vv{{\boldsymbol v}}
\def\pv{{\boldsymbol p}}
\def\av{{\boldsymbol a}}
\def\bv{{\boldsymbol b}}
\def\kv{{\boldsymbol k}}
\def\omnL{\omega_{\mu\nu}}
\def\omnU{\omega^{\mu\nu}}
\def\omnLbar{{\bar \omega}_{\mu\nu}}
\def\omnUbar{{\bar \omega}^{\mu\nu}}
\def\omnLbardot{{\dot {\bar \omega}}_{\mu\nu}}
\def\omnUbardot{{\dot {\bar \omega}}^{\mu\nu}}

\def\oabL{\omega_{\alpha\beta}}
\def\oabU{\omega^{\alpha\beta}}
\def\omnLD{{\tilde \omega}_{\mu\nu}}
\def\omnUD{\tilde {\omega}^{\mu\nu}}
\def\omnLDbar{{\bar {\tilde \omega}}_{\mu\nu}}
\def\omnUDbar{{\bar {\tilde {\omega}}}^{\mu\nu}}

\def\epsLmnbg{\epsilon_{\mu\nu\beta\gamma}}
\def\epsUmnbg{\epsilon^{\mu\nu\beta\gamma}}
\def\epsLmnab{\epsilon_{\mu\nu\alpha\beta}}
\def\epsUmnab{\epsilon^{\mu\nu\alpha\beta}}

\def\epsUmnrs{\epsilon^{\mu\nu\rho \sigma}}
\def\epsUlnrs{\epsilon^{\lambda \nu\rho \sigma}}
\def\epsUlmrs{\epsilon^{\lambda \mu\rho \sigma}}

\def\epsLmnbg{\epsilon_{\mu\nu\beta\gamma}}
\def\epsUmnbg{\epsilon^{\mu\nu\beta\gamma}}
\def\epsLmnab{\epsilon_{\mu\nu\alpha\beta}}
\def\epsUmnab{\epsilon^{\mu\nu\alpha\beta}}

\def\epsLabgd{\epsilon_{\alpha\beta\gamma\delta}}
\def\epsUabgd{\epsilon^{\alpha\beta\gamma\delta}}

\def\epsUmnrs{\epsilon^{\mu\nu\rho \sigma}}
\def\epsUlnrs{\epsilon^{\lambda \nu\rho \sigma}}
\def\epsUlmrs{\epsilon^{\lambda \mu\rho \sigma}}

\def\epsLijk{\epsilon_{ijk}}
\def\half{\frac{1}{2}}

\def\GLW{{\rm GLW}}

\def\n0{n_{(0)}}
\def\e0{\varepsilon_{(0)}}
\def\P0{P_{(0)}}
\title{Theoretical aspects of relativistic spin hydrodynamics coupled with electromagnetic fields}

\author*[a,b]{Rajeev Singh}
\affiliation[a]{Center for Nuclear Theory, Department of Physics and Astronomy, Stony Brook University, Stony Brook, New York, 11794-3800, USA}
\affiliation[b]{Department of Modern Physics, University of Science and Technology of China, Hefei, Anhui 230026, China}
\emailAdd{rajeevofficial24@gmail.com}
\author[c]{Masoud Shokri}
\affiliation[c]{Institute for Theoretical Physics, Goethe University Frankfurt,
Max-von-Laue-Strasse 1, D-60438 Frankfurt am Main, Germany}
\emailAdd{shokri@itp.uni-frankfurt.de}
\author[d]{S.~M.~A.~Tabatabaee Mehr}
\affiliation[d]{School of Particles and Accelerators, Institute for Research in Fundamental Sciences (IPM), P.O. Box 19395-5531, Tehran, Iran}
\emailAdd{tabatabaee@ipm.ir}

\abstract{We expand the classical phase-space distribution function to incorporate couplings between spin and electromagnetic fields. This extension has led to the derivation of modified constitutive relations for the charge current, energy-momentum tensor, and spin tensor. Due to these couplings, the new tensors are revised from their perfect fluid analogues, creating an interplay between the background and spin fluid equations of motion.
The corrections introduced in our framework have the potential to shed light on the experimentally observed discrepancies in the spin polarization measurements of Lambda hyperons.}

\FullConference{%
  25th International Spin Symposium (SPIN 2023)\\
  24-29 September, 2023}

\begin{document}

\def\be{\begin{equation}}
	\def\ee{\end{equation}}
	\def\barr{\begin{array}}
	\def\earr{\end{array}}
\def\beq{\begin{eqnarray}}
	\def\eeq{\end{eqnarray}}
\def\bfig{\begin{figure}}
	\def\efig{\end{figure}}
\newcommand{\bea}{\begin{eqnarray}}
	\newcommand{\eea}{\end{eqnarray}}

\def\LB{\left(}
\def\RB{\right)}
\def\LSB{\left[}
\def\RSB{\right]}
\def\LAB{\langle}
\def\RAB{\rangle}

\newcommand{\VP}{\vphantom{\frac{}{}}\!}
\def\lt{\left}
\def\rt{\right}
\def\CHI{\chi}
\newcommand{\rfmtwo}[2]{Eqs.~(\ref{#1})-(\ref{#2})}
\newcommand{\rfcs}[1]{Refs.~\cite{#1}}

\def\a{\alpha}
\def\b{\beta}
\def\g{\gamma}
\def\d{\delta} 
\def\r{\rho}
\def\s{\sigma}
\def\c{\chi} 
 \def\lam{\lambda} 
\def\LR{\left(} 
\def\RR{\right)}
\def\LS{\left[} 
\def\RS{\right]}
\def\LC{\left{} 
\def\RC{\right}}
\def\LA{\left\langle}
\def\RA{\right\rangle}
\def\LD{\left.}
\def\RD{\right.}
\def\HP{\hphantom{\alpha}} 



\def\half{\frac{1}{2}}

\def\GLW{{\rm GLW}}
\def\LRFF{{\rm LRFF}}


\def\nU{n_{(0)}}
\def\nUi{n_{(0),i}}
\def\eU{\varepsilon_{(0)}}
\def\eUi{\varepsilon_{(0),i}}
\def\PU{P_{(0)}}
\def\PUi{P_{(0),i}}
\def\sU{s_{(0)}}
\def\sU{s_{(0),i}}

\def\nP{n_{}}
\def\eP{\varepsilon_{}}
\def\PP{P_{}}
\def\sP{s_{}}
\def\wP{w_{}}

\def\nn{\nonumber}



\def\cA{{\cal A}}
\def\cB{{\cal B}}
\def\cC{{\cal C}}
\def\cD{{\cal D}}
\def\cN{{\cal N}}
\def\cE{{\cal E}}
\def\cP{{\cal P}}
\def\cS{{\cal S}}
\def\cT{{\cal T}}
\def\cQ{{\cal Q}}
\def\cNN{{\cal N}_{(0)}}
\def\cEN{{\cal E}_{(0)}}
\def\cPN{{\cal P}_{(0)}}
\def\cSN{{\cal S}_{(0)}}


\def\pmu{p^\mu}
\def\pnu{p^\nu}

\def\vv{{\boldsymbol v}}
\def\pv{{\boldsymbol p}}
\def\av{{\boldsymbol a}}
\def\bv{{\boldsymbol b}}
\def\kv{{\boldsymbol k}}
\def\omnL{\omega_{\mu\nu}}
\def\omnU{\omega^{\mu\nu}}
\def\omnLbar{{\bar \omega}_{\mu\nu}}
\def\omnUbar{{\bar \omega}^{\mu\nu}}
\def\omnLbardot{{\dot {\bar \omega}}_{\mu\nu}}
\def\omnUbardot{{\dot {\bar \omega}}^{\mu\nu}}

\def\oabL{\omega_{\alpha\beta}}
\def\oabU{\omega^{\alpha\beta}}
\def\omnLD{{\tilde \omega}_{\mu\nu}}
\def\omnUD{\tilde {\omega}^{\mu\nu}}
\def\omnLDbar{{\bar {\tilde \omega}}_{\mu\nu}}
\def\omnUDbar{{\bar {\tilde {\omega}}}^{\mu\nu}}
\def\CHI{\chi}
\def\bchem{\mu_{\rm B}}
\def\bfug{\xi_{\rm B}}
\def\tfug{\xi}

\def\be{\begin{equation}}
\def\ee{\end{equation}}
\def\ba{\begin{eqnarray}}
\def\ea{\end{eqnarray}}   

\def\a{\alpha}
\def\b{\beta}
\def\g{\gamma}
\def\d{\delta} 
\def\r{\rho}
\def\s{\sigma}
\def\c{\chi}
 
\def\LR{\left(} 
\def\RR{\right)}
\def\LS{\left[} 
\def\RS{\right]}
\def\LC{\left{} 
\def\RC{\right}}
\def\LA{\left\langle}
\def\RA{\right\rangle}
\def\LD{\left.}
\def\RD{\right.}
\def\half{\frac{1}{2}}

\def\GLW{{\rm GLW}}
\def\LRF{{\rm LRF}}


\def\nU{n_{(0)}}
\def\eU{\varepsilon_{(0)}}
\def\PU{P_{(0)}}
\def\sU{s_{(0)}}

\def\nP{n_{}}
\def\eP{\varepsilon_{}}
\def\PP{P_{}}
\def\sP{s_{}}
\def\wP{w_{}}


\def\pmu{p^\mu}
\def\pnu{p^\nu}

\def\vv{{\boldsymbol v}}
\def\pv{{\boldsymbol p}}
\def\av{{\boldsymbol a}}
\def\bv{{\boldsymbol b}}
\def\cv{{\boldsymbol c}}
\def\Cv{{\boldsymbol C}}
\def\kv{{\boldsymbol k}}
\def\piv{{\boldsymbol \pi}}

\def\thetap{\theta_\perp}
\def\omnL{\omega_{\mu\nu}}
\def\omnU{\omega^{\mu\nu}}
\def\omnLbar{{\bar \omega}_{\mu\nu}}
\def\omnUbar{{\bar \omega}^{\mu\nu}}
\def\omnLbardot{{\dot {\bar \omega}}_{\mu\nu}}
\def\omnUbardot{{\dot {\bar \omega}}^{\mu\nu}}

\def\oabL{\omega_{\alpha\beta}}
\def\oabU{\omega^{\alpha\beta}}
\def\omnLD{{\tilde \omega}_{\mu\nu}}
\def\omnUD{\tilde {\omega}^{\mu\nu}}
\def\omnLDbar{{\bar {\tilde \omega}}_{\mu\nu}}
\def\omnUDbar{{\bar {\tilde {\omega}}}^{\mu\nu}}

\def\epsLmnbg{\epsilon_{\mu\nu\beta\gamma}}
\def\epsUmnbg{\epsilon^{\mu\nu\beta\gamma}}
\def\epsLmnab{\epsilon_{\mu\nu\alpha\beta}}
\def\epsUmnab{\epsilon^{\mu\nu\alpha\beta}}

\def\epsUmnrs{\epsilon^{\mu\nu\rho \sigma}}
\def\epsUlnrs{\epsilon^{\lambda \nu\rho \sigma}}
\def\epsUlmrs{\epsilon^{\lambda \mu\rho \sigma}}

\def\epsLmnbg{\epsilon_{\mu\nu\beta\gamma}}
\def\epsUmnbg{\epsilon^{\mu\nu\beta\gamma}}
\def\epsLmnab{\epsilon_{\mu\nu\alpha\beta}}
\def\epsUmnab{\epsilon^{\mu\nu\alpha\beta}}

\def\epsLabgd{\epsilon_{\alpha\beta\gamma\delta}}
\def\epsUabgd{\epsilon^{\alpha\beta\gamma\delta}}

\def\epsUmnrs{\epsilon^{\mu\nu\rho \sigma}}
\def\epsUlnrs{\epsilon^{\lambda \nu\rho \sigma}}
\def\epsUlmrs{\epsilon^{\lambda \mu\rho \sigma}}

\def\epsLijk{\epsilon_{ijk}}


\def\epsLmnbg{\epsilon_{\mu\nu\beta\gamma}}
\def\epsUmnbg{\epsilon^{\mu\nu\beta\gamma}}
\def\epsLmnab{\epsilon_{\mu\nu\alpha\beta}}
\def\epsUmnab{\epsilon^{\mu\nu\alpha\beta}}

\def\epsUmnrs{\epsilon^{\mu\nu\rho \sigma}}
\def\epsUlnrs{\epsilon^{\lambda \nu\rho \sigma}}
\def\epsUlmrs{\epsilon^{\lambda \mu\rho \sigma}}

\def\epsLmnbg{\epsilon_{\mu\nu\beta\gamma}}
\def\epsUmnbg{\epsilon^{\mu\nu\beta\gamma}}
\def\epsLmnab{\epsilon_{\mu\nu\alpha\beta}}
\def\epsUmnab{\epsilon^{\mu\nu\alpha\beta}}

\def\epsLabgd{\epsilon_{\alpha\beta\gamma\delta}}
\def\epsUabgd{\epsilon^{\alpha\beta\gamma\delta}}

\def\epsUmnrs{\epsilon^{\mu\nu\rho \sigma}}
\def\epsUlnrs{\epsilon^{\lambda \nu\rho \sigma}}
\def\epsUlmrs{\epsilon^{\lambda \mu\rho \sigma}}

\def\epsLijk{\epsilon_{ijk}}
\def\half{\frac{1}{2}}
\def\GLW{{\rm GLW}}

\def\n0{n_{(0)}}
\def\e0{\varepsilon_{(0)}}
\def\P0{P_{(0)}}
\newcommand{\redflag}[1]{{\color{red} #1}}
\newcommand{\blueflag}[1]{{\color{blue} #1}}
\newcommand{\checked}[1]{{\color{darkblue} \bf [#1]}}
\newcommand{\Psis}{{\sf \Psi}}
\newcommand{\psis}{{\sf \psi}}
\newcommand{\Psibar}{{\overline \Psi}}
\def\eMf{electromagnetic (EM) }
\def\EMf{Electromagnetic (EM) }
\def\EM{EM }
\def\lRFf{local rest frame (LRF)}
\def\LRFf{Local rest frame (LRF) }
\def\LRF{LRF }
\def\QGPf{Quark gluon plasma (QGP) }
\def\qGPf{Quark gluon plasma (QGP) }
\def\QGP{QGP }
\def\mHDf{magnetohydrodynamic (MHD) }
\def\MHDf{Magnetohydrodynamic (MHD) }
\def\MHD{MHD }
\def\iMHD{iMHD }
\def\HD{Hydrodynamics }
\def\hD{hydrodynamics }
\def\RHD{Relativistic hydrodynamics }
\def\rHD{relativistic hydrodynamics }
\def\rMHDf{relativistic magnetohydrodynamic (RMHD) }
\def\RMHDf{Relativistic magnetohydrodynamic (RMHD) }
\def\RMHD{RMHD }
\def\eOMf{equations of motion (EOM)~}
\def\EOMf{Equations of motion (EOM)~}
\def\EOM{EOM}
\def\fl{\ensuremath{\text{Fluid}}}
\def\lrf{\ensuremath{\text{LRF}}}
\def\BVf{Boltzmann-Vlasov (BV) }
\def\BV{BV\,}
		
\def\rhoLEQ{{\widehat{\rho}}_{\rm \small LEQ}}
\def\rhoGEQ{{\widehat{\rho}}_{\rm \small GEQ}}
		
\def\fplushat{{\hat f}^+}
\def\fminushat{{\hat f}^-}
		
\def\fplusrs{f^+_{rs}}
\def\fplussr{f^+_{sr}}
\def\fplusrsxp{f^+_{rs}(x,p)}
\def\fplussrxp{f^+_{sr}(x,p)}
		
\def\fminusrs{f^-_{rs}}
\def\fminussr{f^-_{sr}}
\def\fminusrsxp{f^-_{rs}(x,p)}
\def\fminussrxp{f^-_{sr}(x,p)}
		
\def\fpmrs{f^\pm_{rs}}
\def\fpmrsxp{f^\pm_{rs}(x,p)}

\def\feqplus{f^+_{eq}}
\def\feqplus{f^+_{eq}}
\def\feqplusxp{f^+_{eq}(x,p)}
\def\feqplusxp{f^+_{eq}(x,p)}
		
\def\feqminus{f^-_{eq}}
\def\feqminus{f^-_{eq}}
\def\feqminusxp{f^-_{eq}(x,p)}
\def\feqminusxp{f^-_{eq}(x,p)}
	
\def\feqpm{f^\pm_{{\rm eq}}}
\def\feqpmxp{f^\pm_{{\rm eq}}(x,p)}
\def\feqpmi{f^\pm_{{\rm eq},i}}
\def\feqpmxpi{f^\pm_{{\rm eq},i}(x,p)}
\def\fpm{f^\pm}
\def\fpmxp{f^\pm(x,p)}
\def\fpmi{f^\pm_i}
\def\fpmxpi{f^\pm_i(x,p)}
\newcommand{\rs}[1]{\textcolor{red}{#1}}
\newcommand{\rrin}[1]{\textcolor{blue}{#1}}
\newcommand{\rrout}[1]{\textcolor{blue}{\sout{#1}}}
\newcommand{\lie}[2]{\pounds_{#1}\,#2}
\newcommand{\rd}{\mathrm{d}}
\def\re{\mathrm{e}}
\def\echarge{\ensuremath{\rho_e}}
\def\cond{\ensuremath{{\sigma_e}}}
\newcommand{\msnote}[1]{\todo[author=Masoud]{#1}}
\newcommand{\msnotei}[1]{\todo[author=Masoud,inline]{#1}}
\newcommand{\explainindetail}[1]{\todo[color=red!40]{#1}}
\newcommand{\insertref}[1]{\todo[color=green!40]{#1}}
\newcommand{\fm}{\rm{\,fm}}
\newcommand{\fmc}{\rm{\,fm/c}}


\def\uv{{\boldsymbol U}}


\def\kbarzero{ {\bar k}^0}
\def\kv{{\boldsymbol k}}
\def\kbarv{{\bar {\boldsymbol k}}}

\def\obarzero{ {\bar \omega}^0}
\def\ov{{\boldsymbol \omega}}
\def\obar{{\bar \omega}}
\def\obarv{{\bar {\boldsymbol \omega}}}

\def\ev{{\boldsymbol e}}
\def\bv{{\boldsymbol b}}
\newcommand{\tT}{\theta_T}
\newcommand{\UD}[1]{\oU{#1}}
\newcommand{\XD}[1]{\oX{#1}}
\newcommand{\YD}[1]{\oY{#1}}
\newcommand{\ZD}[1]{\oZ{#1}}
\newcommand\oU[1]{\ensurestackMath{\stackon[1pt]{#1}{\mkern2mu\bullet}}}
\newcommand\oX[1]{\ensurestackMath{\stackon[1pt]{#1}{\mkern2mu\star}}}
\newcommand\oY[1]{\ensurestackMath{\stackon[1pt]{#1}{\mkern2mu\smwhitestar}}}
\newcommand\oZ[1]{\ensurestackMath{\stackon[1pt]{#1}{\mkern2mu\circ}}}
\def\Aone{{ \cal A}_1 }
\def\Atwo{{ \cal A}_2 }
\def\Athree{{ \cal A}_3 }
\def\Afour{{ \cal A}_4 }
\def\vv{{\boldsymbol v}}
\def\pv{{\boldsymbol p}}

\newcommand{\inv}[1]{\frac{1}{#1}}
\newcommand{\iinv}[1]{1/#1}
\maketitle
\section{Introduction}
\label{sec:introduction}
%
High-energy physics has greatly benefited from RHIC and LHC experiments, providing insights into hot, dense relativistic nuclear matter~\cite{Florkowski:1321594}. These experiments show that colliding nuclei form a system evolving from a non-equilibrium glasma state to a quark-gluon plasma (QGP) phase, which then recombines into hadrons~\cite{Kharzeev:2000ph,Heinz:2001xi,Gyulassy:2004zy,Shuryak:2004cy}. Relativistic hydrodynamics has revealed that QGP acts like nearly perfect fluid droplets~\cite{Romatschke:2007mq, Heinz:2013th,Son:2006em,Schafer:2009dj}, advancing the study of hydrodynamics, especially in off-equilibrium processes~\cite{Florkowski:2017olj}.
Recent research has increasingly focused on spin polarization in relativistic nuclear matter~\cite{Becattini:2020ngo} and its behavior in electromagnetic fields (EM)~\cite{Kharzeev:2013jha,Hattori:2022hyo}. New methods of measuring spin polarization in particles from these collisions have advanced our understanding of QGP~\cite{STAR:2017ckg,STAR:2019erd,ALICE:2019aid,Acharya:2019ryw,ALICE:2021pzu}, leading to a wave of theoretical progress~\cite{Florkowski:2017ruc,Florkowski:2017dyn,Florkowski:2018ahw,Florkowski:2018fap,Florkowski:2019qdp,Singh:2020rht,Singh:2021man,Florkowski:2021wvk,Bhadury:2020puc,Hattori:2019lfp,Fukushima:2020ucl,Li:2020eon,Montenegro:2020paq,Weickgenannt:2020aaf,Garbiso:2020puw,Gallegos:2021bzp,Sheng:2021kfc,Speranza:2020ilk,Bhadury:2021oat}.
Spin polarization in colliding systems aligns with their global angular momentum, mainly due to spin-orbit coupling~\cite{Liang:2004ph}. At the macroscopic level, the near-equilibrium dynamics of the QGP suggest that spin also thermalizes, potentially creating spin polarization through vorticity-spin interactions~\cite{Becattini:2013fla}. This theory is supported by the match between global polarization data and `spin-thermal' models~\cite{Becattini:2016gvu,Karpenko:2016jyx,Pang:2016igs,Xie:2017upb,Becattini:2017gcx,Fu:2020oxj}. However, these models struggle to explain some detailed observables~\cite{Karpenko:2016jyx,Becattini:2017gcx,Florkowski:2019voj}, despite recent theoretical progress~\cite{Becattini:2021suc,Becattini:2021iol,Fu:2021pok,Florkowski:2021xvy}.
The gap between theoretical models and experimental results indicates an incomplete understanding of spin polarization dynamics in heavy-ion collisions. To address this, spin, if thermalized, should be assimilated into hydrodynamic models as are other macroscopic quantities, allowing for a detailed examination of spin dynamics. 
To this end, various approaches have been developed, including thermodynamic equilibrium~\cite{Becattini:2009wh}, perfect fluid spin hydrodynamics~\cite{Florkowski:2017ruc,Florkowski:2017dyn}, entropy current analysis~\cite{Hattori:2019lfp,Fukushima:2020ucl,Li:2020eon,She:2021lhe,Daher:2022xon,Cao:2022aku}, non-local collisions~\cite{Yang:2020hri,Weickgenannt:2020aaf,Weickgenannt:2021cuo,Sheng:2021kfc,Hu:2021pwh,Das:2022azr,Weickgenannt:2022zxs}, kinetic theory for massless fermions~\cite{Stephanov:2012ki,Chen:2014cla,Hidaka:2018ekt,Shi:2020htn}, holography~\cite{Heller:2020hnq,Garbiso:2020puw,Gallegos:2021bzp,Hongo:2021ona,Gallegos:2022jow}, and anomalous hydrodynamics~\cite{Son:2009tf,Kharzeev:2010gr}. Significant progress has been made in developing relativistic hydrodynamics with spin~\cite{Florkowski:2017ruc, Florkowski:2017dyn,Florkowski:2018ahw,Florkowski:2018fap, Florkowski:2019qdp, Singh:2020rht,Singh:2021man, Florkowski:2021wvk}, including its application to dissipative
systems~\cite{Bhadury:2020puc, Bhadury:2020cop}.
In our work~\footnote{Our study employs the mostly-minus Minkowski metric, $g_{\mu\nu}={\rm diag}(+1,-1,-1,-1)$. Consequently, the fluid four-velocity $U^\mu$ satisfies $U^\mu U_\mu = 1$. We use $\Delta^{\mu\nu}=g^{\mu\nu}-U^\mu U^\nu$ to project tensors orthogonal to $U^\mu$. Any tensor $M_{\mu\nu}$ can be split into symmetric $M_{(\mu\nu)}=\frac{1}{2}(M_{\mu\nu}+M_{\nu\mu})$ and asymmetric $M_{[\mu\nu]}=\frac{1}{2}(M_{\mu\nu}-M_{\nu\mu})$ components. The Levi-Civita symbol $\epsilon^{\alpha\beta\gamma\delta}$, being totally antisymmetric, follows $\epsilon^{0123}=-\epsilon_{0123}=1$. Euclidean three-vectors are in bold contrasting with four-vectors. We denote scalar and Frobenius products as $a \cdot b \equiv a^\mu b_\nu$ and $A : B \equiv A^{\mu\nu}B_{\mu\nu}$. The paper uses natural units where $c = \hbar = k_B = 1$, unless otherwise specified.}, we've broadened the spin hydrodynamics framework to include spin and EM field interactions in the phase-space distribution of particles. This has led to modified constitutive relations for baryon charge current, energy-momentum tensor, and spin tensor
and is particularly relevant considering the observed spin polarization difference between $\Lambda$ and $\Bar{\Lambda}$ particles~\cite{STAR:2017ckg, Li:2021zwq}, potentially due to their opposite magnetic moments~\cite{Becattini:2016gvu}. We anticipate that our expanded framework will be instrumental in understanding and explaining this polarization splitting~\cite{Singh:2022ltu,Singh:2022uyy}.
%
\section{Single particle distribution function with spin-EM coupling}
\label{sec:HydroEM}
We start with the phase-space distribution function for classical particles with spin-$1/2$ and mass $m$ with $\omega_{\alpha \beta}(x)$ representing the spin polarization tensor~\cite{Florkowski:2018fap,Bhadury:2020puc}
 \begin{equation}
f^\pm_{0}(x,p,s) = f^\pm_{0}(x,p)\exp[\half\,\omega(x) :  s(p)]\,,
\label{eq:fpm-spin}
 \end{equation}
where $s^{\alpha\beta}(p)$ denotes the intrinsic angular momentum of the particle, written in terms of spin four-vector $s^{\alpha}$ and four-momentum $p^{\alpha}$ as
$s^{\alpha\beta} = (1/m) \epsUabgd p_\gamma s_\delta$. The Jüttner distribution, $f_{0}^{\pm}(x,p)=\exp[-p \cdot \beta(x)\pm\xi(x)]$, includes $\xi(x)$ as the baryon chemical potential $\mu(x)$ to temperature $T(x)$ ratio, $\xi=\mu/T$, and $\beta_{\mu}(x)$ as the fluid four-velocity $U_\mu(x)$ to temperature ratio, $\beta_\mu=U_\mu/T$. It's noted that this classical distribution function, as in Eq.~\eqref{eq:fpm-spin}, is initially applicable for local particle collisions but can be adapted for non-local effects via gradients of $f^\pm_{0}(x,p,s)$~\cite{Florkowski:2018fap}.
Generalizing Eq.~\rfn{eq:fpm-spin} to include a coupling between magnetic moment of the particle and external EM field takes the form
\begin{equation}
	f^\pm_{\rm s} (x,p,s) = f^\pm_{\rm 0}(x,p,s)\exp\left[\mp\alpha_M(x) F(x) : s\right],
	\label{eq:fpm-em-spin}
\end{equation}
with $F_{\mu\nu}$ being the Faraday tensor written in terms of $E_\mu$ (electric) and $B_\mu$ (magnetic) four-vectors.
In \rf{eq:fpm-em-spin}, $\alpha_M$ is defined as $\mu_{M}/T$, where $\mu_{M} = g_{Q} \mu_{N}$ represents the magnetic moment of the quasiparticles, with $\mu_{N}$ being the nuclear magneton. For simplicity, this work assumes quasiparticles as $\Lambda$ hyperons, having $g_\Lambda = -0.6138 \pm 0.0047$~\cite{ParticleDataGroup:2020ssz}. However, a more realistic model would include multiple quark-like quasiparticles with distinct masses~\cite{Singh:2021man}. 
Considering the low amplitude of spin polarization observed in measurements~\cite{STAR:2017ckg}, we adopt the small polarization limit, $\omega_{\alpha\beta} \ll 1$~\cite{Florkowski:2019qdp}. We also assume weak EM fields, where $eB \ll M^2$ and $M$ is the typical energy scale of the system. For electrically neutral $\Lambda$ hyperons, $M$ approximates to $m_\Lambda$. These assumptions allow for a simplified approximation of \rf{eq:fpm-em-spin} as
\begin{eqnarray}
		f^\pm_{\rm s}(x,p,s) &=& f_{0}^{\pm}(x,p)\left(1 +\f{1}{2}  \omega : s\right)
	\left(1 \mp \FUGM F : s\right),
		~~~~\label{feq}
	\end{eqnarray}
with $f_{\rm s}^+(f_{\rm s}^-)$ being the particle (antiparticle) distribution function. 
Using the fluid velocity four-vector $\omega_{\alpha \beta}$ can be decomposed as~\cite{Florkowski:2021wvk}
 \begin{equation}
\omega_{\mu\nu} = \kappa_\mu U_\nu - \kappa_\nu U_\mu + \epsilon_{\mu\nu\a\b} U^\a \omega^{\b}, \quad \text{where} \quad \kappa_\mu= \omega_{\mu\a} U^\a, \quad \omega_\mu = \half \epsilon_{\mu\a\b\gamma} \omega^{\a\b} U^\gamma, \label{spinpol1}
 \end{equation}
are orthogonal to $U^\mu$~\cite{Florkowski:2021wvk}.
The constraints on $\kappa_\mu$ and $\omega_\mu$ restrict them to three degrees of freedom each, equating to the number of degrees of freedom in $\omega_{\mu\nu}$. These four-vectors are expressible using three orthonormal space-like vectors $X^\mu$, $Y^\mu$, and $Z^\mu$, which, along with $U^\mu$, constitute a basis in the plane orthogonal to $U^\mu$. 
These vectors satisfy the relations $U \cdot U = 1$ and $X\cdot X = Y \cdot Y = Z \cdot Z = -1$ forming a complete basis set~\cite{Florkowski:2019qdp,Florkowski:2021wvk}. Consequently,
 \begin{align}
\kappa^\a =  C_{\kappa X} X^\a + C_{\kappa Y} Y^\a + C_{\kappa Z} Z^\a\,, \quad
\omega^\a =  C_{\omega X} X^\a + C_{\omega Y} Y^\a + C_{\omega Z} Z^\a\,,
\lab{eq:o_decom}
 \end{align}
where $C_{\boldsymbol{\kappa}} = (C_{\kappa X}, C_{\kappa Y}, C_{\kappa Z})$, and
$C_{\boldsymbol{\omega}} = (C_{\omega X}, C_{\omega Y}, C_{\omega Z})$ are spin polarization components.
\section{Constitutive relations}
\label{sec:cons_relations}
Within the presence of the coupling between the spin and EM fields, we now derive the constitutive relations using Eq.~\eqref{feq} \cite{Singh:2022ltu}.

The first constitutive relation is the baryon charge current
defined as the first moment of Eq.~(\ref{feq})~\cite{Florkowski:2018fap,Florkowski:2018ahw}
 \begin{equation}
N^\lambda 
= \!\int \! \mathrm{dP}~\mathrm{dS} \, \, p^\lambda \, \left[f^+_{\rm s}(x,p,s)\!-\!f^-_{\rm s}(x,p,s) \right],
\label{eq:Neq-sp0}
 \end{equation}
where momentum $\mathrm{dP}$ and spin $\mathrm{dS}$ integration measures are defined, respectively, as~\cite{Florkowski:2018fap}
 \begin{align}
\mathrm{dP} = \frac{\mathrm{d}^3p}{(2 \pi )^3 E_p}\,,\quad \mathrm{dS} = \frac{m}{\pi {\mathfrak{s}}} \, \mathrm{d}^4s~\delta(s\cdot s + {{\mathfrak{s}}}^2)~\delta(p\cdot s)\,,
\label{eq:dS}
 \end{align}
with $E_p$ and $\mathfrak{s}^2$ being the particle energy and length of the spin vector, respectively.
In analogy to the dissipative charged fluid current~\cite{Kovtun:2012rj}, $N^\lambda$ is written as
 \begin{equation}
N^{\lambda} =  {\cal N} U^{\lam} + N_\perp^\lam = \left({\cal N}_{\rm PF}+{\cal N}_{\rm EM}\right) U^{\lam} + N_\perp^\lam \,,
\label{eq:chargecurrent}
 \end{equation}
with $\cN_{\rm PF}$ being the baryon charge density for ideal fluid~\cite{Florkowski:2021wvk}
\ba
{\cal N}_{\rm PF} = 4 \, \sinh(\xi)\, {\cal N}_{(0)}\,.
\lab{nden}
\ea
The term $4 \, \sinh(\xi)$ accounts for both spin and particle-antiparticle degeneracies, while ${\cal N}_{(0)}$ represents the number density of spinless, neutral classical massive particles~\cite{Florkowski:1321594}
 \begin{equation} 
{\cal N}_{(0)} = gT^3\,z^2   K_{2}(z)\,, \label{N0}
 \end{equation}
with $g = 1/(2\pi^2)$,
$z=m/T$ and $K_n$ being $n^{th}$
modified Bessel function of 2$^{nd}$ kind.
The coupling between spin and EM fields give rise to ${\cal N}_{\rm EM}$ and the transverse current $N_\perp^\lam$
\ba
{\cal N}_{\rm EM} &=& \alpha_M  \cosh(\xi){\cal N}_{(0)} \epsilon^{\beta\gamma\nu\mu} \omega_{\beta\gamma} U_\mu B_\nu \,,\nonumber\\
N_\perp^\lam &=& \alpha_M  \cosh(\xi) \mathcal{A}_3 \big(U^\lam F^{\beta\gamma} + 6 U^\lam U^{[\beta} E^{\gamma]} - U^{[\beta} F^{\gamma]\lam} - g^{\lambda
[\beta} E^{\gamma]}\big)\omega_{\beta\gamma}\,,
\label{eq:couplingNden}
\ea
with
$
{\cal A}_3 = -\left(2\left({\cal E}_{(0)} + {\cal P}_{(0)}\right)\right)/(T\, z^2)\, 
$~\cite{Florkowski:2017ruc,Florkowski:2018fap,Florkowski:2018ahw,Florkowski:2021wvk}, and ${\cal P}_{(0)}$ and ${\cal E}_{(0)}$ being the pressure, and energy density expressed as~\cite{Florkowski:1321594}
 \begin{equation}
{\cal P}_{(0)} = {\cal N}_{(0)} T\,, \qquad
{\cal E}_{(0)} =  g\, z^3 T^4 K_{1}(z) + 3 {\cal P}_{(0)}\,, \label{P0}
 \end{equation}
respectively. 
From the conservation of charge current, $\partial_\mu N^\mu = 0$, we have
 \begin{equation}
\UD {\mathcal{N}}_{\rm PF} + \UD{\mathcal{N}}_{\rm EM} + \left(\mathcal{N}_{\rm PF} + \mathcal{N}_{\rm EM}\right) \theta_U  = - \partial \cdot N_\perp\,,
\label{eq:NEoMgeneralMS}
 \end{equation}
where $\UD {\cdots}\equiv U \cdot\p\,{\cdots}$ is the comoving temporal derivative with $\theta_U\equiv\p \cdot U$ being the expansion scalar. 
The electric current is expressed as
\begin{equation}
J^\mu ~=~ q \, N^\mu\,,
\label{eq:J}
\end{equation}
for quasiparticles having an electric charge $q$.
This results in the back-reaction of spin-EM coupling with EM fields through Maxwell equations \cite{Singh:2022ltu}.
As we assume Lambda hyperons to be electrically neutral quasiparticles of the fluid the electric current tends to zero~\cite{Denicol:2019iyh}.

The second constitutive relation is the energy-momentum tensor defined as~\cite{Florkowski:2018fap,Florkowski:2018ahw}
\begin{eqnarray}
T^{\mu \nu}
&=& \int \mathrm{dP}~\mathrm{dS} \, \, p^\mu p^\nu \, \left[f^+_{\rm s}(x,p,s) + f^-_{\rm s}(x,p,s) \right].
\label{eq:Teq-sp02}
\end{eqnarray}
Plugging Eq.~\eqref{feq} we obtain
 \begin{equation}
T^{\mu \nu} = \cE U^\mu U^\nu - \cP \Delta^{\mu\nu} + \cQ^\mu U^\nu+ \cQ^\nu U^\mu +\cT^{\mu\nu}\,,
\label{eq:TMU1}
 \end{equation}
with modified energy density
\begin{eqnarray}
\cE &\equiv& U_\mu U_\nu T^{\mu \nu} =
\mathcal{E}_{\rm PF} + \mathcal{E}_{\rm EM}\,,
\label{eq:cE}
\end{eqnarray}
where
 \begin{align}
{\cal E}_{\rm PF} = 4 \cosh(\xi) {\cal E}_{(0)}, \quad
{\cal E}_{\rm EM} = \alpha_M
\sinh(\xi)\bigg\{ {\cal E}_{(0)} \,\omega : F + 2 \Big[\left(\Inqr{4}{0}{0}+\Inqr{4}{1}{0}\right)\kappa\cdot E -2\Inqr{4}{1}{0}\omega\cdot B\Big]\bigg\}\,,\label{eq:edensity}
 \end{align}
and modified pressure
\begin{eqnarray}
\cP &\equiv& -\frac{1}{3}\Delta : T =  \mathcal{P}_{\rm PF} + \mathcal{P}_{\rm EM}\,,
			\label{eq:cP}
\end{eqnarray}
with
 \begin{align}
{\cal P}_{\rm PF} = 4 \cosh(\xi)  {\cal P}_{(0)}, \quad
{\cal P}_{\rm EM} = \alpha_M
\sinh(\xi) \left\{{\cal P}_{(0)} \omega : F 
- 2 \left[\left(\Inqr{4}{1}{0}+\frac{5}{3}\Inqr{4}{2}{0}\right)\kappa\cdot E
-\frac{10}{3}\Inqr{4}{2}{0}\omega\cdot B\right]\right\}.\label{eq:pressure}
 \end{align}
$\cQ^\mu$ represents transverse vector current in \rf{eq:TMU1}. This is similar to heat current, whereas $\cT^{\mu\nu}$ denotes transverse traceless tensor which resembles stress tensor for the case of dissipative fluid~\cite{Kovtun:2012rj}
\begin{eqnarray}
			\cQ^\mu &\equiv& \Delta^{\mu}_{\HP\a}U_\b T^{\a\b}= 2\,\FUGM \sinh(\FUG) \Inqr{4}{1}{0}\epsUmnab\,U_\nu\left(E_\alpha \omega_\beta-B_\alpha \kappa_\beta \right),
			 \label{eq:cQ}
			 \\
			 \cT^{\mu\nu} &\equiv& \Delta^{\mu\nu}_{\alpha\beta} \, T^{\a\b}
			 = 
			 4\,\FUGM \sinh(\FUG)\Inqr{4}{2}{0}\Big(
			 E^{(\mu}\kappa^{\nu)}+ B^{(\mu}\omega^{\nu)}-
		 \inv{3}\Delta^{\mu\nu}\left(\kappa\cdot E+\omega \cdot B\right)
			 \Big)\,.
			 \label{eq:cT}
\end{eqnarray}
where $\Delta^{\mu\nu}_{\alpha\beta} \equiv (1/2)\left[\Delta^\mu_{\HP\a}\Delta^\nu_{\HP\b}+\Delta^\nu_{\HP\a}\Delta^\mu_{\HP\b}-(2/3)\Delta^{\mu\nu}\Delta_{\alpha\beta}\right]$.
Energy-momentum conservation, $\partial_\nu T^{\mu\nu} = F^{\mu\sigma} J_\sigma$, can be written in two parts. The first part (energy equation) is longitudinal to the fluid flow \cite{Singh:2022ltu}
\begin{eqnarray}
\UD{\cE} + (\cE + \cP) \, \theta_U ~=~ - q E \cdot {N_\bot } + \UD{U} \cdot \mathcal{Q}  -  \nabla \cdot\mathcal{Q}  
+ \half\,\mathcal{T}_{\mu\nu} \sigma^{\mu\nu}  \,,
\label{eq:energyEq}
\end{eqnarray}
where $\nabla_\mu \equiv \partial_\mu - U_\mu U^\nu \partial_\nu$ and
$\s^{\mu \nu }\equiv \Delta^{\mu\nu}_{\alpha\beta} \nabla^\alpha U^\beta$,
while the second part is transverse to  $U^\mu$~\cite{Singh:2022ltu}
\begin{eqnarray}
(\mathcal{E} + \mathcal{P}) \, \UD{U}^{\mu }  &=& \nabla^{\mu } \mathcal{P}+q\left(\cN E^{\mu } + \epsilon^{\mu\alpha \beta \sigma } B_{\alpha }{N_{\bot}}_\beta \, U_{\sigma } \right)- \Big( \UD{U} \cdot \mathcal{Q}  +  \frac{1}{2}\,
\mathcal{T}_{\alpha\beta} \, \sigma^{\alpha\beta} \Big) U^{\mu }
\nn\\
&-& 
\p_{\alpha }\mathcal{T}^{\mu \alpha } -  2 \UD{\cQ}^{\mu } - \theta_U \cQ^\mu \,.
\label{eq:eulerEqn}
\end{eqnarray}
Finally, the spin tensor is expressed as~\cite{Florkowski:2018fap}
 \begin{equation}
S^{\lambda, \mu\nu} = \int  \mathrm{dP}~\mathrm{dS} \, \, p^\lambda \, s^{\mu \nu} 
\left[f^+_{\rm s}(x,p,s) + f^-_{\rm s}(x,p,s) \right],
\label{SGLW}
 \end{equation}
which, after using Eq.~\eqref{feq}, gives~\cite{Singh:2022ltu}
 \begin{equation}
S^{\lambda, \mu\nu} = S^{\lambda, \mu\nu}_{\rm PF} - 2 \alpha_M \tanh(\xi) S^{\lambda, \mu\nu}_{\rm EM}\,,
\label{SGLW1}
 \end{equation}
where
 \begin{align}
S^{\a,\b\g}_{\rm PF}
&= \cosh(\xi) \Big[{\cal A}_1 \, U^\a \,  \omega^{\b\g} 
+ {\cal A}_2 \, U^\a \,  U^{[\b} \, \omega^{\g]}_{\HP\d} \, U^\d +  {\cal A}_3 \, \left(U^{[\b} \, \omega^{\g]\a}  + g^{\a[\b}\, \omega^{\g]}_{\HP\d} \, U^\d \right)\Big]\,,
\nonumber\\
S^{\a,\b\g}_{\rm EM}
&= \cosh(\xi)\Big[ U^\a  {\cal A}_1\,  F^{\b\g} 
+ {\cal A}_2 \,U^\a U^{[\b} \, F^{\g]}_{\HP\HP\d} \, U^\d  +  {\cal A}_3 \, \left(U^{[\b} \, F^{\g]\a}  + g^{\a[\b}\, F^{\g]}_{\HP\HP\d} \, U^\d \right)\Big],
\label{SEM2}
 \end{align}
and thermodynamic coefficients are: ${\cal A}_1 =  {\cal N}_{(0)}-{\cal A}_3$,
${\cal A}_2 = 2 \left[{\cal A}_1 -2 {\cal A}_3\right]$.
The phase-space distribution function, Eq.~\eqref{feq}, applicable to local particle collisions, indicates that spin $(S^{\alpha,\beta\gamma})$ can be conserved independently of the orbital angular momentum~\cite{Singh:2022ltu}.
In this study, the energy-momentum tensor~\eqref{eq:Teq-sp02} is symmetric by definition, but could include antisymmetric contributions from non-local collisional effects~\cite{Weickgenannt:2022zxs,Das:2022azr}, which are omitted here for simplicity. Consequently, disregarding non-local collisions, the spin tensor acts as a conserved current
\begin{equation}
    \p_\alpha\, S^{\alpha,\beta\gamma} = 0\,,
    \label{eq:SPIN}
\end{equation}
allowing us to obtain six equations of motion for spin polarization components.
\section{Summary}
\label{sec:summary}
In this study, we have innovatively incorporated a spin-EM coupling term into the classical phase-space distribution function, leading to a redefined framework for calculating the charge current, energy-momentum tensor, and spin tensor. This approach is a significant deviation from earlier models~\cite{Florkowski:2019qdp,Singh:2022uyy}, primarily due to the intricate coupling between the background and spin equations of motion introduced by the spin-EM interaction.
This formalism holds promise for explaining the experimentally observed spin polarization splitting between $\Lambda$ and $\Bar{\Lambda}$ particles. To fully realize its potential, it is imperative to incorporate realistic conditions into the model. Future research will be directed towards these more complex scenarios, aiming to bridge the gap between theoretical predictions and experimental observations in the realm of high-energy heavy-ion collisions.

\smallskip
{\it Acknowledgements.}
R.S. acknowledges the support of Polish NAWA Bekker program No.: BPN/BEK/2021/1/00342 and  Polish National Science Centre Grant No. 2018/30/E/ST2/00432.
M.S. acknowledges support by the Deutsche Forschungsgemeinschaft (DFG, German Research Foundation) through the CRC-TR 211 `Strong-interaction matter under extreme conditions'– project number 315477589 – TRR 211.
\bibliographystyle{utphys}
\bibliography{fluctuationRef.bib}{}
\end{document}